\documentclass[aps,prb,twocolumn,floatfix]{revtex4-2}
\usepackage{graphicx}
\usepackage[version=4]{mhchem}
\usepackage[hidelinks]{hyperref}
\usepackage{bm}
\makeatletter

\newcommand{\Rmnum}[1]{\expandafter\@slowromancap\romannumeral #1@}

\usepackage{amsmath}   
\usepackage{amssymb}   

\usepackage{siunitx,physics} 
\DeclareSIUnit{\angstrom}{\text{\AA}}

\usepackage{color}

\makeatother
\begin{document}
\title{Exciton fine structure in CdSe nanoplatelets using a quasi‑2D screened configuration‑interaction framework}
\author{Sumanti Patra}
\email{sumanti1357@gmail.com}
\address{University of Hamburg, Institute of Physical Chemistry, Luruper Chaussee 149, 22761 Hamburg}
\address{The Hamburg Centre for Ultrafast Imaging, Hamburg, Germany}
\author{Gabriel Bester}
\email{gabriel.bester@uni-hamburg.de}
\address{University of Hamburg, Institute of Physical Chemistry, Luruper Chaussee 149, 22761 Hamburg}
\address{The Hamburg Centre for Ultrafast Imaging, Hamburg, Germany}

\begin{abstract}
We compute exciton binding energies and fine‑structure splittings in CdSe nanoplatelets with two zincblende geometries and one wurtzite geometry, finding that the wurtzite structure exhibits the largest bright‑bright splitting due to its intrinsic in‑plane anisotropy, while the zincblende structures show smaller but finite splittings arising from atomistic symmetry breaking at edges and corners. These results are obtained using a theoretical framework that we developed, which combines DFT single‑particle states with screened configuration interaction, a quasi‑2D dielectric screening model, and an efficient Coulomb‑cutoff scheme that eliminates periodic‑image interactions and enables accurate Coulomb and exchange integrals at low computational cost. This methodology provides a transferable and practical route for studying excitons in CdSe nanoplatelets and other quasi‑two‑dimensional nanomaterials.
\end{abstract}	
\maketitle

\section{Introduction}
Two-dimensional nanoplatelets (NPL) have drawn significant attention due to their tunable optical characteristics, sharp photoluminescence peaks, large absorption cross-section and high quantum efficiency~\cite{ithurria2011colloidal,yu2020optical,diroll20232d, bertram2006nanoplatelets, niebur2023untangling, li20172d, cherusseri2020synthesis, cui2021efficient, khan2017near, roda2023colloidal, tessier2012spectroscopy}, making them ideal for advanced optoelectronic devices such as light-emitting diodes~\cite{cirignano2025blue,khan2020cdse} and photodetectors~\cite{dutta20222d}. Due to their atomic-scale thickness, they show very strong quantum confinement in one direction. NPLs are now synthesized from a wide range of materials, including chalcogenides~\cite{ithurria2011colloidal,yu2020optical,diroll20232d,roda2023colloidal,khan2017near}, transition metal dichalcogenides (TMDCs)~\cite{bertram2006nanoplatelets, niebur2023untangling}, perovskites~\cite{li20172d, cherusseri2020synthesis, cui2021efficient}, each offering unique properties and advantages. In particular, II–VI semiconductor nanoplatelets have emerged as one of the strongest candidates for the narrowest room-temperature emitters due to their high exciton binding energy~\cite{diroll20232d} and precise control over thickness and lateral dimensions~\cite{ithurria2008quasi}. Their strong binding energy also enables stabilization of both neutral and charged excitons (trions)\cite{ayari2020tuning}. Among these, CdSe nanoplatelets~\cite{diroll20232d, ithurria2011colloidal, cui2021efficient,tessier2012spectroscopy,yu2020optical} have been most widely studied and synthesized and are well known for their excellent optical properties in the visible range.  
CdSe NPLs can be synthesized in both zincblende~\cite{bouet2013two,diroll20232d} and wurtzite crystal structures~\cite{kong2023covalent,gao2018distinct}, and this provides additional tunability in their excitonic properties.

Preliminary theoretical work includes the screened configuration‑interaction (scrCI) methodology, originally developed for quantum dots \cite{franceschetti1999many, bester2008electronic,luo2009long} and more recently extended to periodic 2D materials \cite{torche2019first}. However, NPLs represent a  challenge to these approaches because their large lateral dimensions make \textit{ab initio} atomistic RPA screening, as used in \cite{torche2019first}, computationally prohibitive. Conversely, the quantum dot approach is only well suited for bulk‑like nanostructures, where reliable bulk screening models exist \cite{resta1977thomas} (also refined to include confinement effects in a simplified manner  \cite{penn1962wave, karpulevich2019dielectric}). This quantum‑dot‑based approach, where the structures are rather large, can also benefit from a good quality semi‑empirical description of the electronic structure using empirical pseudopotentials \cite{bester2008electronic, bui2020excitonic} or tight binding \cite{bryant2003tight,schulz2005tight,koskinen2003tight} which is not available for 2D materials. 
The scrCI methodology has the advantage to accurately captures electron–hole Coulomb and exchange interactions as well as correlation effects and naturally leads to all excitations, such as trions\cite{morch2025beyond}, biexcitons \cite{torche2021biexcitons} etc..., which is an advantageous feature we aim to exploit for an accurate description of NPLs.

In the present work, we use density functional theory (DFT) for computing single-particle states. This has the advantage to lead to a rather accurate single-particle description and gives a full flexibility in terms of the choice of materials, but brings along the disadvantage of a methodology that is limited by the capability of DFT codes in terms of system sizes. For the NPLs the approach is useful as we can reach the experimentally realized sizes, although on the low size range. 
We extend the methodology for describing electron–hole Coulomb and exchange interactions to two‑dimensional NPLs, where strong out‑of‑plane and weaker in‑plane confinement shape the excitonic states. To capture the anisotropic screening of these quasi‑2D systems, we incorporate a 2D dielectric screening model, in contrast to the isotropic treatment appropriate for 0D quantum dots. A key challenge is the accurate evaluation of long‑range Coulomb integrals without introducing spurious interactions with periodic images. Enlarging the simulation cell is both computationally prohibitive and ultimately ineffective, so we adopt a simple real‑space cutoff scheme with separate vertical and lateral cutoff functions tailored to the NPL geometry. Despite its simplicity, this approach yields highly accurate Coulomb integrals at minimal computational cost, ensuring that integral evaluation is not a bottleneck. We demonstrate the general framework using CdSe NPLs and analyze their characteristic optical properties.

\section{Computational Methods} 
The first step involves solving the single particle problem H$\psi_{\alpha}$ = $E_{\alpha}$ $\psi_{\alpha}$ and calculating the single-particle states $\psi_{i\alpha}$ corresponding to energy $E_{\alpha}$ for the two-dimensional nanoplatelets using density functional theory (DFT). Based on these states, we then construct a set of single-substitution Slater determinants {$\Phi_{v,c}$} by putting one electron from an occupied valence state  with energy $E_v$ to an unoccupied conduction state with energy $E_c$, starting from the ground-state Slater determinant {$\Phi_{0}$}. Each exciton wavefunction $\Psi$(m), characterized by the exciton quantum numbers $ m $ , is constructed as a superposition of Slater determinants from the chosen basis set.
\begin{eqnarray}
\Psi^{(m)} = \sum_{v=1}^{N_v} \sum_{c=1}^{N_c} A_{v,c}^{(m)} \, \Phi_{v,c}
\end{eqnarray}
where N$_v$ and N$_c$ are the number of valence band and conduction band respectively included in the basis of the exciton wavefunctions.

The matrix element of the many-body Hamiltonian $ H^{MB}$ has been calculated in the
basis $\{\Phi_{v c}\}$
\begin{multline}
\langle \Phi_{v_1c_1} | H^{MB} | \Phi_{v_2 c_2}  \rangle = (E_{c_1} - E_{v_1}) \delta_{v_1,v_2} \delta_{c_1,c_2} -J_{v_1c_1v_2c_2} \\+ 
 K_{v_1c_1v_2c_2}
\end{multline}
where $J_{v_1c_1v_2c_2}$ and $K_{v_1c_1v_2c_2}$ are the Coulomb and Exchange matrix elements given as: 
\begin{multline}
J_{v_1c_1v_2c_2}
\label{eq:Jvcv'c'}
= \sum_{s_1,s_2} \iint 
\psi^*_{v_1}(\mathbf{r_1}, s_1)\, \psi^*_{c_2}(\mathbf{r_2}, s_2)\, 
W(\mathbf{r_1}, \mathbf{r_2})\, \\
\times \psi_{v_2}(\mathbf{r_1}, s_1)\, \psi_{c_1}(\mathbf{r_2}, s_2)\, 
d\mathbf{r_1}\, d\mathbf{r_2}
\end{multline}
\begin{multline}
K_{v_1c_1v_2c_2}
= \sum_{s_1,s_2} \iint 
\psi^*_{v_1}(\mathbf{r_1}, s_1)\, \psi^*_{c_2}(\mathbf{r_2}, s_2)\, 
W(\mathbf{r_1}, \mathbf{r_2})\, \\
\times \psi_{c_1}(\mathbf{r_1}, s_1)\, \psi_{v_2}(\mathbf{r_2}, s_2)\, 
d\mathbf{r_1}\, d\mathbf{r_2}
\end{multline}
and $W(\mathbf{r_1}, \mathbf{r_2})$ is the screened Coulomb interaction that has the general form:
\begin{eqnarray}
W(\mathbf{r_1}, \mathbf{r_2}) = e^2 \int \varepsilon^{-1}(\mathbf{r_1}, \mathbf{r_3})\hspace{1mm}|\mathbf{r_3}-\mathbf{r_2}|^{-1} \hspace{1mm}d\mathbf{r_3} \quad .
\end{eqnarray}
The Coulomb matrix element given in equation (\ref{eq:Jvcv'c'}) can be rewritten in terms of the electrostatic potential $\phi$ : 
\begin{multline}
\label{}
J_{v_1c_1v_2c_2} = \sum_{s_1} \int 
\psi^*_{v_1}(\mathbf{r}_1, s_1)\, 
\psi_{v_2}(\mathbf{r}_1, s_1)\,\phi_{c_2c_1}(\mathbf{r}_1)\, d\mathbf{r_1}
\label{eq:J_eqn2}
\end{multline}
where the electrostatic potential $\phi$ is given as: 
\begin{eqnarray}
    \phi_{c_2c_1}(\mathbf{r_1}) & = & \sum_{s_2} \int W(\mathbf{r_1}, \mathbf{r_2})\, \psi^*_{c_2}(\mathbf{r_2}, s_2)\, \psi_{c_1}(\mathbf{r_2}, s_2)\, d\mathbf{r_2} \\
    & = & \int W(\mathbf{r_1}, \mathbf{r_2})\,\rho_{c_2c_1}(\mathbf{r_2} )\,d\mathbf{r_2}
\end{eqnarray}
where $\rho_{c_2c_1}(\mathbf{r_2} )$ = $\sum_{s_2} \psi^*_{c_2} (\mathbf{r_2}, s_2)\, \psi_{c_1}(\mathbf{r_2}, s_2)$ plays the role of a charge density and is generally a complex function.
The potential $\phi_{c_2c_1}(\mathbf{r_1})$ is calculated in reciprocal space using the convolution theorem. 
\begin{equation}
    \phi_{c_2c_1}(\mathbf{r_1}) = \sum_{\mathbf{G}}  W(\mathbf{G})\, \rho_{c_2c_1} (\mathbf{G}) \, e^{i\mathbf{G}\cdot\mathbf{r_1}} \quad .
    \label{eq:phi_in_g}
\end{equation}
This reciprocal-space approach is advantageous as it significancy reduces computational costs compared to direct real-space computations. It is also convenient for the implementation of various functional forms of the screening function, which are often expressed in momentum space. In this context, the function $ W(\mathbf{G}) $ in equation~(\ref{eq:phi_in_g}) represents the Fourier transform of the screened Coulomb interaction.  This screening 
$ W(\mathbf{G}) $ is calculated by an explicit model $ W( \mathbf{q})$ as a function of the momentum transfer vector $ \mathbf{q} $ .

Our objective is to calculate the Coulomb integrals for 2D nanoplatelets accurately, taking into account their anisotropic dielectric environment. For two dimensional layered materials,  the screened Coulomb potential $ W( \mathbf{q})$ is defined as:
\begin{eqnarray}
\label{eq:Wq_form01}
W( \mathbf{q}) = \frac{4 \pi}{\varepsilon_{2D}(\mathbf{q}_{\|}) q^2}  = \frac{4 \pi}{\varepsilon_{2D}(\mathbf{q}_{\|}) \hspace{1mm} (\mathbf{q}_{\|}^2 + q_z^2)}
\label{Wq_in_cylindrical}
\end{eqnarray}
Here, $\mathbf{q}_{\|}$ and $q_z$ denote the in-plane and out-of-plane components of the momentum transfer vector 
$\mathbf{q}$, respectively. In cylindrical polar coordinate, we can write $\mathbf{q}$ as: \begin{equation} 
\label{eq:qcylindrical}
\mathbf{q}  =  \mathbf{q}_{\|} \hat{q}_{\|} + q_z \hat{z} 
=  q_{\|} \cos\phi \hat{x} + q_{\|} sin\phi \hat{y} + q_z \hat{z} 
\end{equation}
and $\varepsilon_{\text{2D}}(\mathbf{q}_{\|})$
is the effective macroscopic dielectric function for the quasi-2D system. 
Several models can be used to describe the in-plane dielectric screening in 2D materials, such as the Quantum Electrostatic Heterostructure (QEH) model~\cite{andersen2015dielectric} or the analytical model of Trolle \textit{et al.}~\cite{trolle2017model}. While QEH is well suited for van der Waals heterostructures, for CdSe we employ the Trolle model~\cite{trolle2017model} to capture the quasi-2D dielectric response.
This model provides a macroscopic dielectric function for an isotropic quasi-2D material (only the length of the vector $\mathbf{q_{\|}}$ enters the equation as $q_{\|}$) and effectively bridges 3D and 2D descriptions by incorporating a screening function derived from a 3D formulation into the quasi-2D dielectric response. The expression for this quasi-2D dielectric function is given as 
\begin{eqnarray}
&&\varepsilon_\mathrm{2D}(q_{\|}) = \frac{\varepsilon_\mathrm{3D}(q_{\|})}{q_{\|}d} \times \\ \nonumber
&&\hspace{2mm}\frac{\left[1 + \varepsilon_\mathrm{3D}(q_{\|}) + (1 - \varepsilon_\mathrm{3D}(q_{\|})) e^{-q_{\|}d} \right] [q_{\|}d - 1 + e^{-q_{\|}d}]}{\left[1 + \varepsilon_\mathrm{3D}(q_{\|}) + (1 - \varepsilon_\mathrm{3D}(q_{\|})) e^{-q_{\|}d} \right] + 2(e^{-q_{\|}d} - 1)} 
\label{eq:trolle}
\end{eqnarray}

where $d$ is the thickness of the layer, defined as the vertical distance between the lowest and the topmost Cd atoms, and $\varepsilon_\mathrm{3D}(q_{\|})$ is 
a dielectric function that originates from a 3D screening model. Here, we have evaluated $\varepsilon_\mathrm{3D}(q_{\|})$ using the 3D dielectric screening model of Resta~\cite{resta1977thomas}, which describes dielectric screening in bulk materials as a function of the length $q$ of the 3D momentum transfer vector $\mathbf{q}$:
\begin{equation}
    \varepsilon_\mathrm{3D}(q) = \frac{q^2 + q_\mathrm{TF}^2} {q^2 + \frac{q_\mathrm{TF}^2 }{\epsilon_{0}} \frac{\sin(q r_{\infty})}{q r_{\infty}}}\quad ,
\end{equation}
where $q_{TF}$ is the Thomas-Fermi wave-vector, $\epsilon_{0}$ is the static dielectric constant, and  $r_{\infty}$ is the screening radius of the bulk material.

\begin{figure}[h]
\centering
\includegraphics[width=0.9\linewidth]{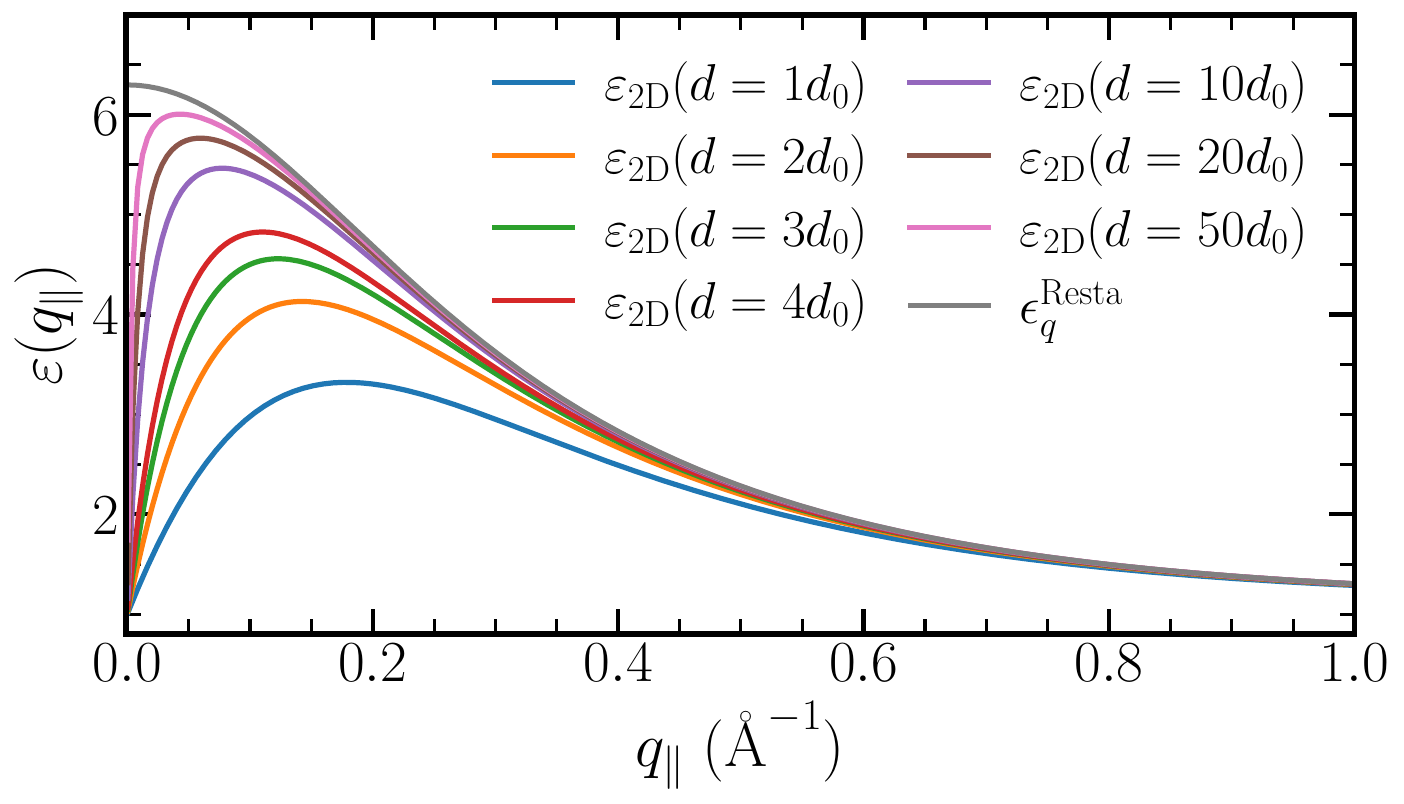}
\caption{
Quasi-2D dielectric function $ \varepsilon_{2D}(q_{\parallel}) $ of CdSe 
calculated using the Trolle \textit{et al.}~\cite{trolle2017model} model along the 
3D screening function of Resta~\cite{resta1977thomas}(blue line) used in the Trolle model. 
Results are shown for several effective layer thicknesses $ d = L d_0 $ 
($ L = 1, 2, 3, 4, 10, 20, 50 $ ), where $ d_0 $ is 3.1 \AA. 
As the layer thickness increases, the quasi-2D dielectric response gradually approaches 
the 3D limit.
}
\label{fig:eps2D}
\end{figure}

Figure~\ref{fig:eps2D} shows the calculated quasi-2D dielectric function 
$ \varepsilon_{2D}(q_{\|}) $ for CdSe using the parameters 
$q_\mathrm{TF}$ = \SI{1.05137}{\bohr^{-1}}, 
$\epsilon_0$ = 6.3, and 
$r_{\infty}$ = \SI{3.703161}{\bohr}. 
Results are presented for several layer thicknesses $ d = L d_0 $ , 
where the thickness  $d_0$ = \SI{3.10}{\angstrom} is defined as the out-of-plane distance between successive Cd atomic sublayers.

As we use periodic boundary conditions in the calculation of the Coulomb and exchange matrix elements, we need to eliminate the electrostatic interaction between periodic images of the nanoplatelets, both in the in-plane and out-of-plane (z) directions, For this purpose, a cutoff scheme specifically designed for 2D geometries is required. To apply this cut-off effectively, the screened Coulomb potential needs to be Fourier transformed to real space: 
\begin{eqnarray}
         W(\mathbf{r}) = \frac{1}{(2 \pi)^3} \int
         W(\mathbf{q})  e^{i\mathbf{q}\cdot\mathbf{r}}  d\mathbf{q}\hspace{32 mm} \nonumber \\
         \label{eq:FT_wqtoWr}
         =\frac{1}{(2 \pi)^3}\int_{q_{\|}=0}^{\infty} \int_{\phi=-\pi}^{\pi} \int_{q_z=-\infty}^{\infty} W(\mathbf{q}) e^{i\mathbf{q}\cdot\mathbf{r}} q_{\|}  dq_{\|}\hspace{0.2mm} d\phi dq_z \quad ,
\end{eqnarray}
where we have used cylindrical coordinates, where  $q_{\|}$ is the in plane component, $\phi$ is the azimuthal angle and $q_z$ is the vertical component of the momentum vector $\mathbf{q}$. The momentum vector  $\mathbf{q}$ is expressed according to Eq.~(\ref{eq:qcylindrical}); similarly, $\mathbf{r} =  \mathbf{r}_{\|}\hat{r_{\|}} + r_z \hspace{1mm} \hat{z} =  r_{\|} \cos\theta \hspace{1mm} \hat{x} + r_{\|} sin\theta \hspace{1mm} \hat{y} + r_z \hspace{1mm} \hat{z}$ and the dot product becomes $\mathbf{q}\cdot\mathbf{r} = (q_{\|} r_{\|} \cos(\phi-\theta)+ q_z z)$ and we can rewrite equation (\ref{eq:FT_wqtoWr}) as:
\begin{multline}
\label{eq:FT_wqtoWr2}
         W(r_{\|},\theta,z) = \frac{1}{(2 \pi)^3}\int_{q_{\|}=0}^{\infty} \int_{\phi=-\pi}^{\pi} \int_{{q_z}=-\infty}^{\infty} W({q_{\|},\phi,q_z}) \\ 
         \hspace{0.5cm} e^{i[q_{\|} r_{\|} \cos(\phi-\theta)+ q_z z]}  \hspace{1mm} q_{\|} dq_{\|} d\phi dq_z  \quad .
\end{multline}
As from equation (\ref{eq:Wq_form01}), $W(\mathbf{q})$ does not depend on the angle $\phi$ since our model dielectric constant is isotropic in-plane (Eq.~(\ref{eq:trolle})), and we can write $W(\mathbf{q})$ as $W({q_{\|},q_z})$ and the expression becomes:
\begin{multline}
         \label{eq:FT_wqtoWr3}
          W(r_{\|},\theta,z)      
          = \frac{1}{(2 \pi)^3}\int_{q_{\|}=0}^{\infty} \int_{q_z=-\infty}^{\infty} W({q_{\|},q_z}) \hspace{1mm} e^{i q_z z} q_{\|}  \\ 
         \times  \Bigl[  \int_{\phi=-\pi}^{\pi}  e^{iq_{\|} r_{\|} \cos(\phi-\theta)}   \hspace{2mm} d\phi  \Bigr]  \hspace{2mm} dq_{\|} dq_z \quad .
\end{multline}
It is straightforward to show by a variable transformation that the $\theta$ dependence in the last integral drops out. Hence, equation (\ref{eq:FT_wqtoWr3}) takes the form:
\begin{multline*}
          W(r_{\|},z)  = \frac{1}{(2 \pi)^3}\int_{q_{\|}=0}^{\infty} \int_{q_z=-\infty}^{\infty} W({q_{\|},q_z}) \hspace{1mm} e^{i q_z z}  \hspace{2mm} q_{\|} \\ 
          \times \Bigl[  \int_{\phi=-\pi}^{\pi}  e^{iq_{\|} r_{\|} \cos(\phi)} d\phi\Bigr] \hspace{2mm}  dq_{\|}\hspace{1mm} dq_z   \quad ,
\end{multline*}
and
\begin{multline}
\label{eq:FT_wqtoWr4}
W(r_{\|}, z) = \frac{1}{(2\pi)^3} \int_{q_{\|}=0}^{\infty} \int_{q_z=-\infty}^{\infty}
W(q_{\|}, q_z) \, e^{i q_z z} \, q_{\|} \\
\times \Bigl[ 2\pi J_0(q_{\|} r_{\|})  \Bigr] \, dq_{\|} \, dq_z  \quad .
\end{multline}
where $J_0(q_{\|} r_{\|})$ is the  zeroth-order Bessel Function of the first kind, and the expression simplies to:

\begin{multline}
\label{eq:FT_wqtoWr5}
W(r_{\|},z) = \frac{2}{(2 \pi)^2} 
\int_{q_{\|}=0}^{\infty} \int_{q_z=0}^{\infty}
q_{\|} J_0(q_{\|}r_{\|}) \, W(q_{\|},q_z) \\ \times  \cos(q_z z) \, dq_{\|}  \, dq_z  \quad .
\end{multline}

\begin{figure}[h!]
    \centering
    \includegraphics[width=0.85\linewidth]{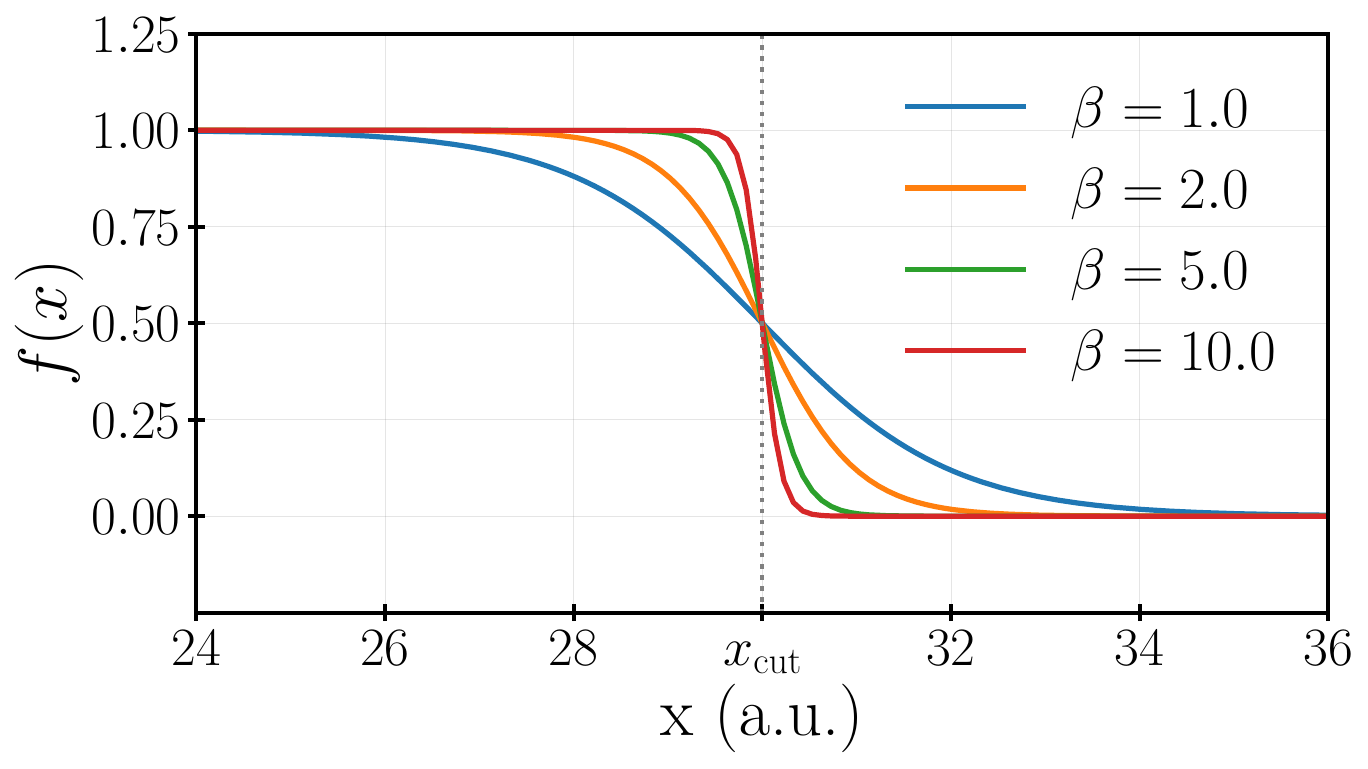}
    \caption{Cut-off function  
    $f(x) = \frac{1}{1 + \exp[\beta(x - x_{\mathrm{cut}})]}$ shown for different values of the smoothness parameter $\beta$ and for a $x_{\mathrm{cut}} = 30 a.u.$. The same functional form applies to both in-plane and out-of-plane directions.}
    \label{fig:cutoff}
\end{figure}

This form of the Coulomb potential $W(r_{\|},z)$  is then truncated in real space using a cut-off function $f_\mathrm{cut}(r_{\|},z)$, in order to eliminate interactions between nanoplatelets located in neighboring periodic supercells. The function  $f_\mathrm{cut}(r_{\|},z)$  is expressed in terms of two different cut-off functions as 
$f_\mathrm{cut}(r_{\|},z)$ =  $f_\mathrm{cut}^\mathrm{in}(r_{\|}) . f_\mathrm{cut}^\mathrm{out}(z)$ where $f_\mathrm{cut}^\mathrm{in}(r_{\|})$ and $f_\mathrm{cut}^\mathrm  {out}(z)$ are the in-plane and 
out-of-plane cut-off functions respectively. These are defined as:
\begin{eqnarray}
f_\mathrm{cut}^\mathrm{in}(r_{\|}) = \frac{1}{1+ \exp{({\beta_\mathrm{in}(r_{\|}-r_{\|\mathrm{cut}}}))}}\quad,
\end{eqnarray} 
\begin{eqnarray}
    f_\mathrm{cut}^\mathrm{out}(z) = \frac{1}{1+ \exp{({\beta_\mathrm{out}(z-z_\mathrm{cut}}))}}  \quad.
\end{eqnarray}
Here, $r_{\|\mathrm{cut}}$ and $z_\mathrm{cut}$ define the radial (in-plane) and vertical (out-of-plane) cut-off distances beyond which the Coulomb interaction is suppressed. The parameters $\beta_\mathrm{in}$ and $\beta_\mathrm{out}$ 
control the smoothness of the cut-off in the respective direction, with larger values leading to a more abrupt truncation.
Since both $f_\mathrm{cut}^\mathrm{in}(r_{\|})$ and $f_\mathrm{cut}^\mathrm{out}(z)$ share the same functional form, we illustrate their behavior using a single representative plot in Fig.~\ref{fig:cutoff}, showing the effect of varying the smoothness parameter $\beta$.
The truncated Coulomb potential is then expressed as $W_\mathrm{cut}(r_{\|},z) = W(r_{\|},z)  f_\mathrm{cut}^\mathrm{in}(r_{\|})  f_\mathrm{cut}^\mathrm{out}(z)$ where we use the assumption that $W_\mathrm{cut}(r_{\|},z)$ has mirror symmetry [= $W_\mathrm{cut}(r_{\|},-z)$ in the derivation of Eq.~(\ref{eq:FT_wqtoWr5}) so that the interaction is effectively cut at $z_\mathrm{cut}$ and $-z_\mathrm{cut}$. 
Then we proceed to transform it into reciprocal space as $W_{\mathrm{cut}}(q_{\|},q_z)$ in cylindrical coordinates. 
The general form of the three-dimensional Fourier transform in cylindrical coordinates is  as 
\begin{eqnarray}
         W(\mathbf{q}) =  \int
         W(\mathbf{r})  e^{-i\mathbf{q}.\mathbf{r}}  d\mathbf{r}\hspace{32 mm} \nonumber \\
         \label{eq:FT_wqtoWq}
         =\int_{r_{\|}=0}^{\infty} \mspace{-2mu}\int_{\theta=-\pi}^{\pi} \mspace{-2mu}\int_{z=-\infty}^{\infty} \mspace{-12mu}W(\mathbf{r}) e^{-i\mathbf{q}\cdot\mathbf{r}} r_{\|}  dr_{\|}\hspace{0.2mm} d\theta dz \quad.
\end{eqnarray}
Exploiting cylindrical symmetry and performing the angular integration as before, we obtain:
\begin{multline}
\mspace{-15mu}W(q_{\|},q_z) = 4\pi \mspace{-6mu}\int_{0}^{\infty}\mspace{-13mu}\int_{0}^{\infty} \mspace{-12mu}r_{\|}\, J_0(q_{\|} r_{\|})\, W(r_{\|},z)
\cos(q_z z)\, dz\, dr_{\|}
\label{eq:wcut_qspace_00}
\end{multline}

\noindent This transformation is then applied to the truncated potential $ W_{\mathrm{cut}}(r_{\|}, z) $ , resulting in:
\begin{multline}
W_{\mathrm{cut}}(q_{\|},q_z) = 4\pi \int_{0}^{\infty} \mspace{-12mu}\int_{0}^{\infty} r_{\|} \, J_0(q_{\|} r_{\|}) \, W(r_{\|},z) \,  f_{\mathrm{cut}}^\mathrm{in}(r_{\|}) \\
\times f_{\mathrm{cut}}^\mathrm{out}(z) \, \cos(q_z z) \, dz \, dr_{\|}
\label{eq:wcut_qspace_01}
\end{multline}

\section{Results and Discussion}
Once we established our methodology, we went on applying it to investigate the excitons in  two-dimensional CdSe nanoplatelets. CdSe nanoplatelets are available in both zincblende (ZB) and wurtzite (WZ) crystal structures.  Their growth direction is influenced by the synthesis environment, and a few specific orientations are typically observed. Experimentally, it has been found that CdSe nanoplatelets can be synthesized with thicknesses as low as three monolayers~\cite{ji2020dielectric}, providing an excellent platform for investigating quantum confinement in quasi–two-dimensional systems.

\begin{figure}[h!]
    \centering
    \includegraphics[width=0.82\linewidth]{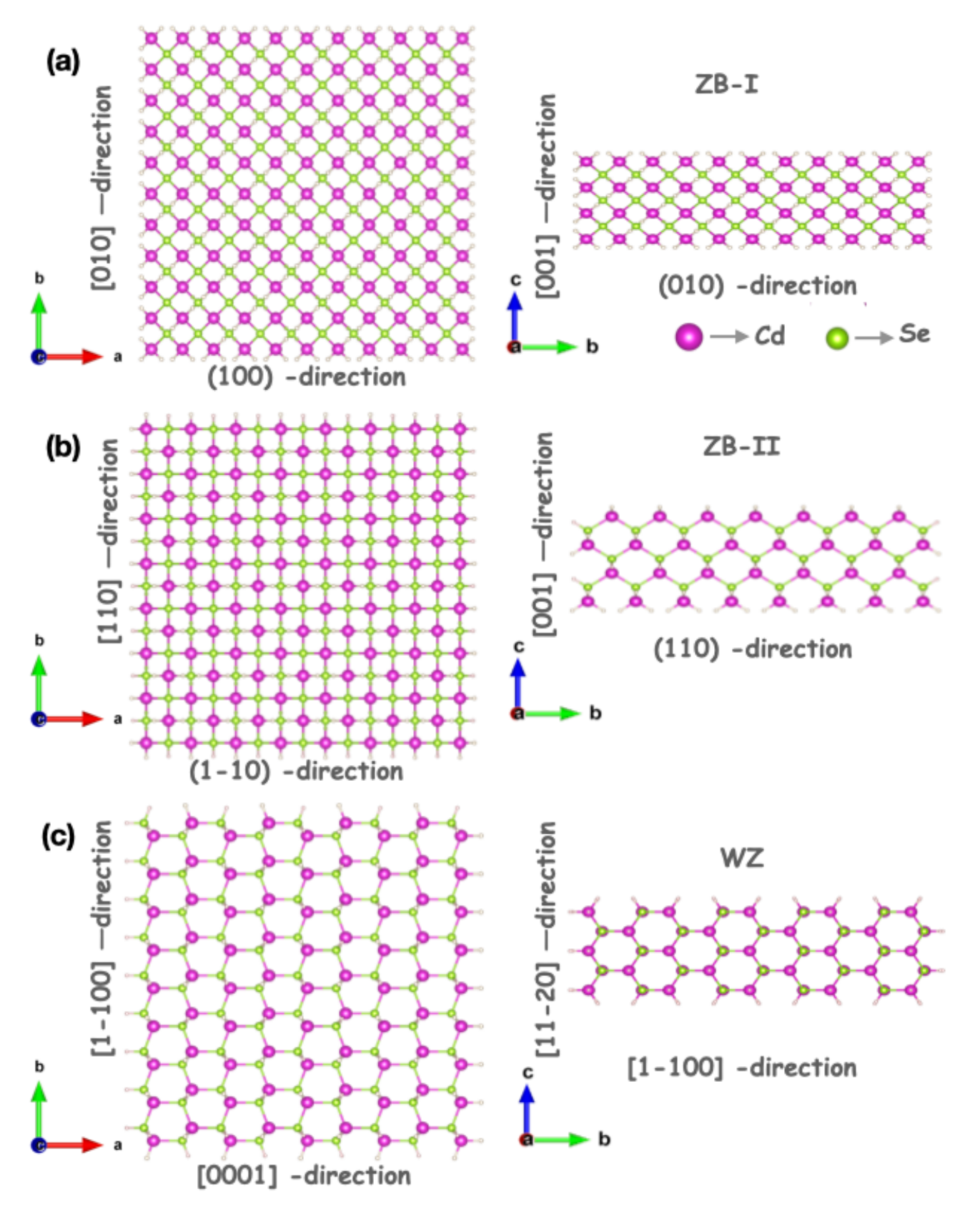}
    \caption{CdSe nanoplatelet structure for three different configurations : (a) ZB(I) (b) ZB(II)  and (c) WZ}
    \label{fig:cdse_npl_structure}
\end{figure}

The three nanoplatelet structures considered in our calculations, ZB(I), ZB(II) and WZ, are shown in Figure~\ref{fig:cdse_npl_structure}. 
For the zincblende (ZB) phase, the most commonly synthesized CdSe nanoplatelets are non-stoichiometric, typically terminated with cadmium (Cd) layers on both the top and bottom surfaces~\cite{bouet2013two,shornikova2018addressing}.
In zincblende nanoplatelets, the growth direction is along [001], with side facets most frequently aligned along either [100] and [010] referred as (ZB(I)) shown in Figure~\ref{fig:cdse_npl_structure}a) or [110] and [1$\bar{1}$0] referred as (ZB(II)) shown in Figure~\ref{fig:cdse_npl_structure}b)~\cite{bouet2013two}. In contrast, for the wurtzite (WZ) phase, the [11$\bar{2}$0] direction is commonly observed as the growth direction, with side facets aligned along [11$\bar{0}$0] and [0001] (WZ, Figure~\ref{fig:cdse_npl_structure}c)~\cite{wang2015two} 


We have considered nanoplatelets with varying lateral sizes and a fixed thickness corresponding to three layers of CdSe.
The ZB(I), ZB(II), and WZ nanoplatelets have thicknesses of 9.30 {\AA}, 9.30 {\AA}, and 8.77 {\AA}, respectively. The dangling bonds at the edges have been passivated with pseudo-hydrogens that carry a fractional charge of 1.5e and 0.5e when connected to Cd and Se, respectively. This type of passivation has been used extensively to simulate a defect-free structure  ~\cite{li2005band,zhang1996method,zhang2016pseudo}.

\begin{figure}[h!]
    \centering
    \includegraphics[width=0.99\linewidth]{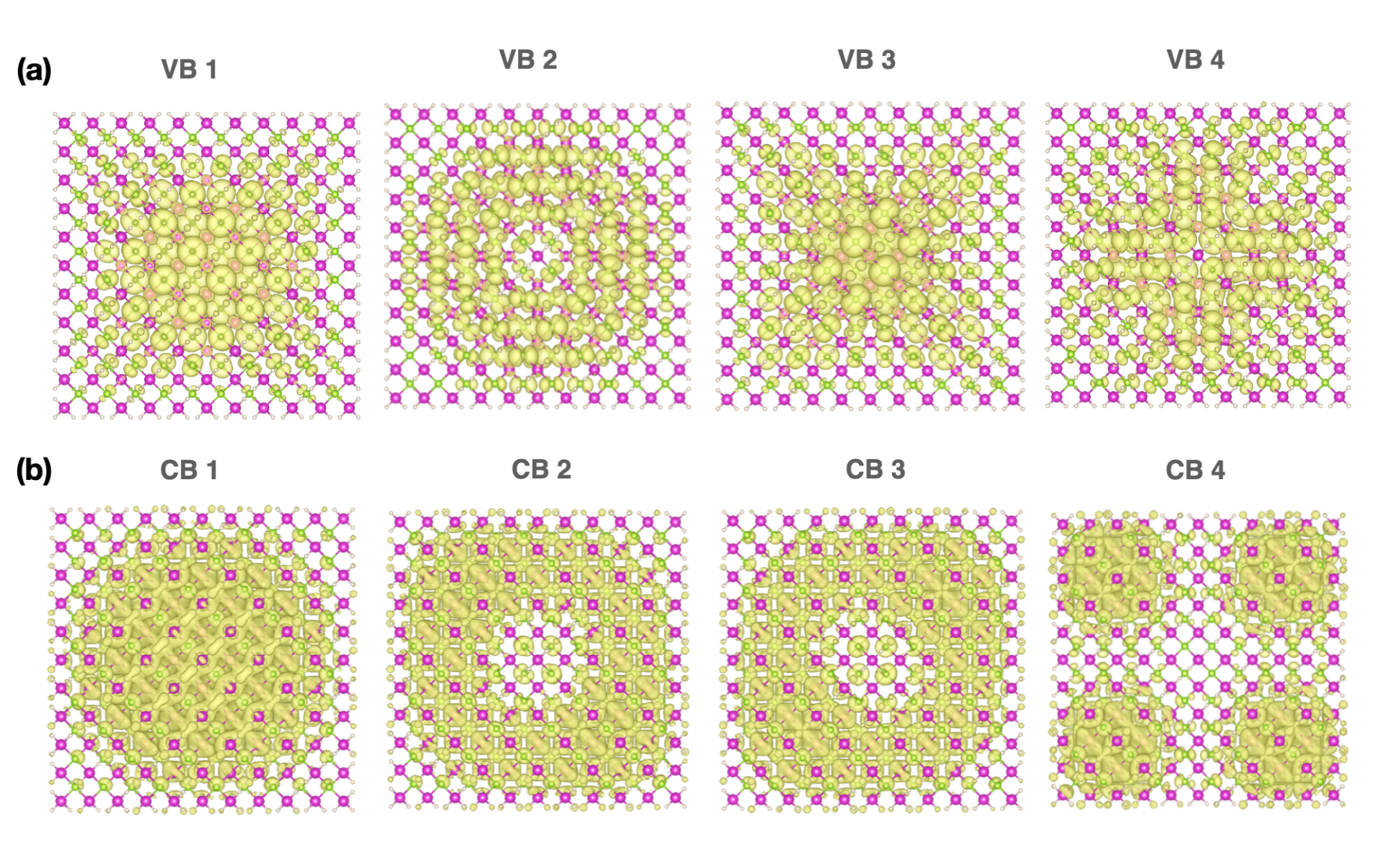}
    \caption{ Four topmost valence band states (a) and bottom most conduction band states (b) for the zincblende (I) structure with a square lateral dimension of $ 31 \times 31\,\text{\AA}^2$ 
    }
    \label{fig:cdse_zb100_wfn}
\end{figure}

To illustrate the methodology in detail, the approach is first applied to a 31 × 31 \AA$^2$ ZB(I) type nanoplatelet with a total number of 750 atoms, which is treated as a representative case for the analysis. In order to calculate the Coulomb and exchange matrix elements, the single-particle wavefunctions $ \psi^*_{i\alpha}(\mathbf{r}_1, s_1)$ are expressed on a real-space grid and obtained from the \textsc{Quantum ESPRESSO} package with norm-conserving pseudopotentials, including spin-orbit coupling. The energetically highest (lowest)   four state of the valence (conduction) band are shown in figure \ref{fig:cdse_zb100_wfn}. The highest occupied valence band state is labeled as VB 1 and the lowest unoccupied conduction band state as CB 1. The splitting between the valence band states is: 
[$\Delta E_{\text{VB}}(12), \Delta E_{\text{VB}}(23), \Delta E_{\text{VB}}(34)$] = [45, 22, 23 meV] and between the conduction band states: [$\Delta E_{\text{CB}}(12), \Delta E_{\text{CB}}(23), \Delta E_{\text{CB}}(34)$] = [203, 5, 281 meV].
The wavefunction look very similar to the well studied case of self-assembled quantum dots~\cite{lei2010artificial}, or even more similar to thickness fluctuation quantum dots~\cite{luo2009atomistic} where the electron state look like a simple single-band object with S, P and D overall orbital character while the valence band holes are multiband and more complex.

\begin{figure}[h!]
    \centering
    \includegraphics[width=0.98\linewidth]{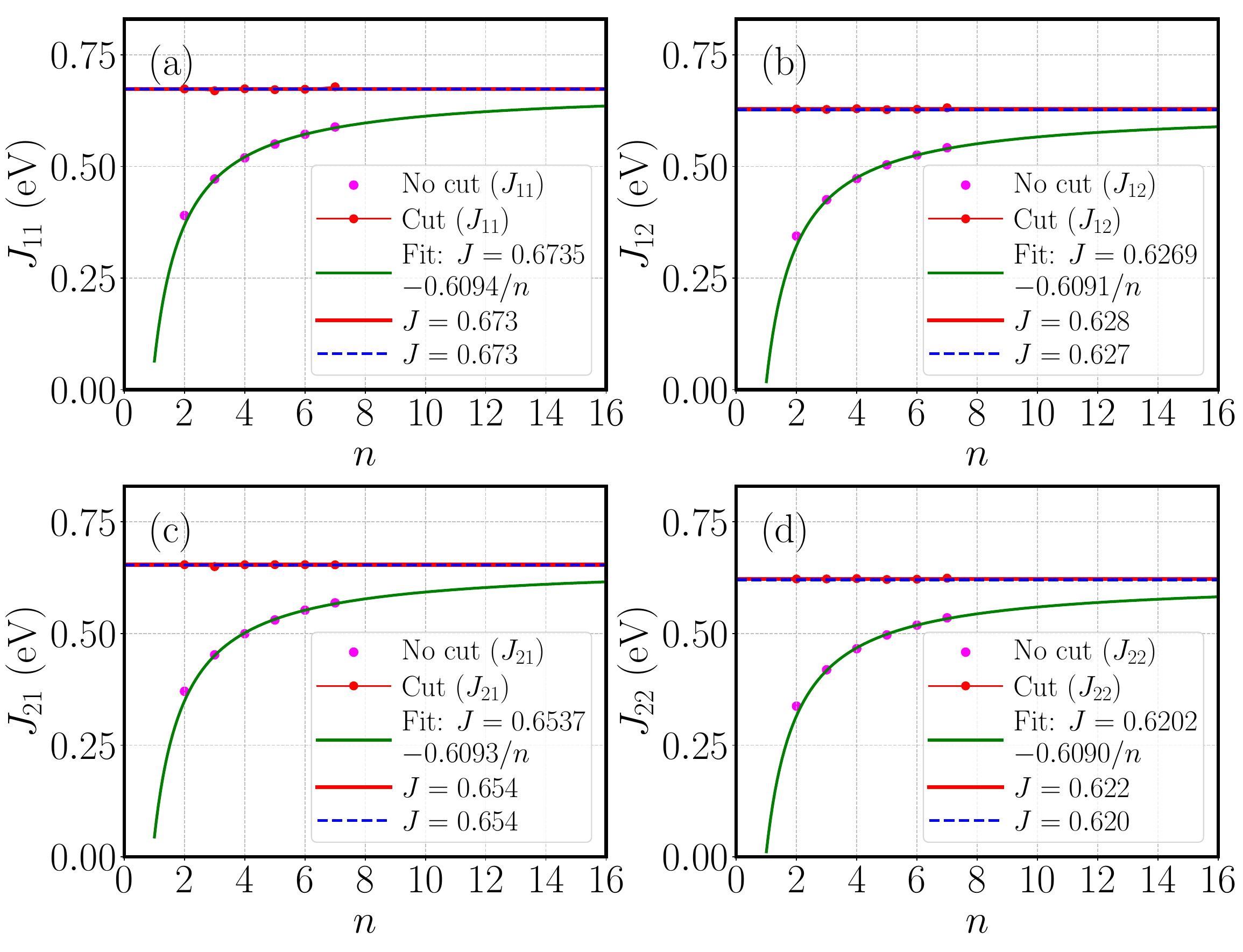}
    \caption{
Coulomb integrals $J_{\mathrm{vc}}$: (a) $J_{11}$, (b) $J_{12}$, (c) $J_{21}$, and (d) $J_{22}$ as a function of $n$, where the supercell has size $n \times n \times n$ and larger $n$ increases the vacuum separation between periodic images. Magenta dots show the raw data without the Coulomb cut-off. The green curve is a fit to $J(n) = a + b/n^{2}$, with $a$, the converged value at infinite vacuum separation, shown by the blue dashed line. Results from the Coulomb cut-off method are shown as red dashed lines.  
}
    \label{fig:j11_j12_j21_j22_vs_n_cdse100}
\end{figure}

To account for the long-range nature of the Coulomb interaction and eliminate spurious periodic-image effects, we employed cut-off functions that truncate the interaction in both in-plane and out-of-plane directions, as detailed in the Methods section. Smoothness parameters $\beta_\mathrm{in}$ and $\beta_\mathrm{out}$ were both set to 2, which gives converged Coulomb integrals with the cut-off method.
Figure~\ref{fig:j11_j12_j21_j22_vs_n_cdse100} shows the Coulomb matrix elements $ J_{\mathrm{vc}}$ = $ J_{\mathrm{vc},\mathrm{vc}}$ for CdSe ZB(I) nanoplatelet as a function of vacuum separation. Here, $ v$ and $ c$ denote the valence band and conduction band indices, and the subplots correspond to the matrix elements $ J_{11}, J_{12}, J_{21},$ and $ J_{22}$ . The results obtained using our Coulomb cut-off approach are shown as red lines, while calculations without applying the cut-off are shown as magenta points. The calculations were performed for different vacuum separations by constructing supercells of size $ n \times n \times n $ , where $ n $ is an integer multiple of the original DFT cell (with a lateral size of 50~\AA ~ and a vacuum spacing of 25~\AA) . Increasing $ n $ effectively enlarges the vacuum region around the 2D nanoplatelets, thereby reducing the interactions between their periodic images. 
The uncut data (magenta) were fitted using the function $ J(n) = a + b/n^2 $ , where $ a $ represents the converged value at infinite vacuum separation. The fitted curve is shown as a green solid line, while the extrapolated value of $ a $ is shown as a blue dashed line. Using this comparison, we validated our reciprocal-space Coulomb cut-off method. The Coulomb integrals obtained from the cut-off method show excellent agreement with the extrapolated reference values, differing by less than 2\%.

After establishing the methodology to obtain accurate Coulomb and exchange integrals at a manageable numerical cost we proceed to use these integrals into a screened configuration interaction scheme ~\cite{franceschetti1999many,bester2003pseudopotential}. In this approach, the exciton many-body wavefunction is represented as a linear combination of Slater determinants in the electron-hole picture, where the Hilbert space is made of 16 conduction band states (that can be occupied by the electron) and 16 valence band states (that can be occupied by the hole). At the exciton level, this approach is equivalent to the Bethe Salpeter method, but using our 2D model screening (instead of a typical RPA screening), implemented as described above.

\begin{figure}[h!]
    \centering
    \includegraphics[width=0.85\linewidth]{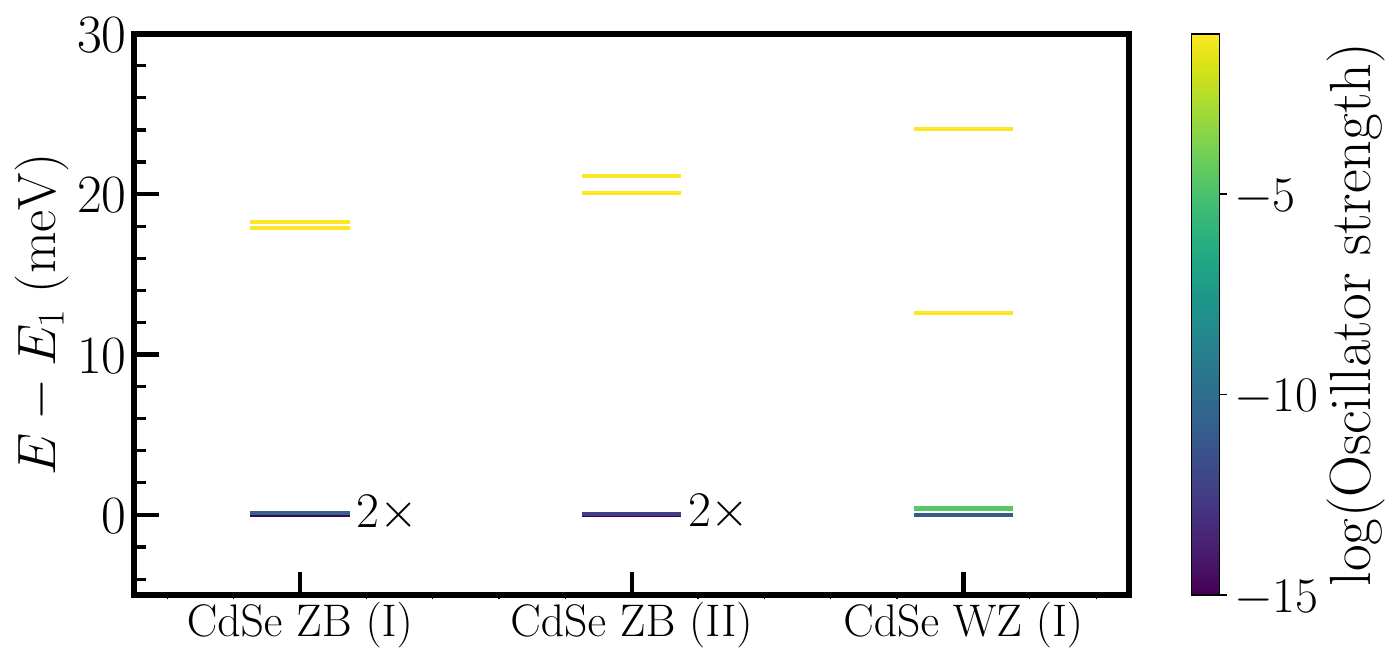}
    \caption{Comparison of excitonic fine structure splitting (FSS) for three types of CdSe nanoplatelets: ZB(I), ZB(II), and WZ(I) with a square lateral dimension of $ 31 \times 31\,\text{\AA}^2$ .  
        Log of the Oscillator strength is plotted in the color code. The lowest dark excitonic state is taken as the reference energy level $ E_1$ .    
        }
    \label{3case_FSS_compare01}
\end{figure}

The results are shown in Fig.~\ref{3case_FSS_compare01} for the three CdSe nanoplatelet types considered in this study: (i) ZB(I), (ii) ZB(II), and (iii) WZ(I). The excitonic fine structure splitting (FSS) was computed for each structure using the screened configuration interaction approach described above. Figure~\ref{3case_FSS_compare01} compares the resulting FSS values across the three geometries.
In all three cases studied, the lowest excitonic states are optically dark. The optically active bright excitonic states lie only a few meV above the dark states. For the ZB(I) structure, the lowest bright excitonic state is found to be 18\,meV above the dark state, and the bright states themselves are nearly degenerate, with a fine splitting of approximately 358\,\textmu eV. In the case of the ZB(II) structure, the bright excitonic states are separated from the lowest dark state by just 20\,meV, and these bright states are also closely spaced, with a splitting of about 1\,meV. In contrast, for the WZ(I) nanoplatelet, the bright excitonic states lie 11\,meV above the dark state but are more strongly split among themselves, with an energy difference of about 12\,meV. This strong splitting among the bright states in WZ(I) is attributed to the lack of in-plane symmetry between two lateral $ x$ and $ y$ directions in the Wurtzite crystal structure. This analysis was carried out for few distinct lateral dimensions for each type of nanoplatelets.

\begin{figure}[h!]
    \centering
    \includegraphics[width=0.98\linewidth]{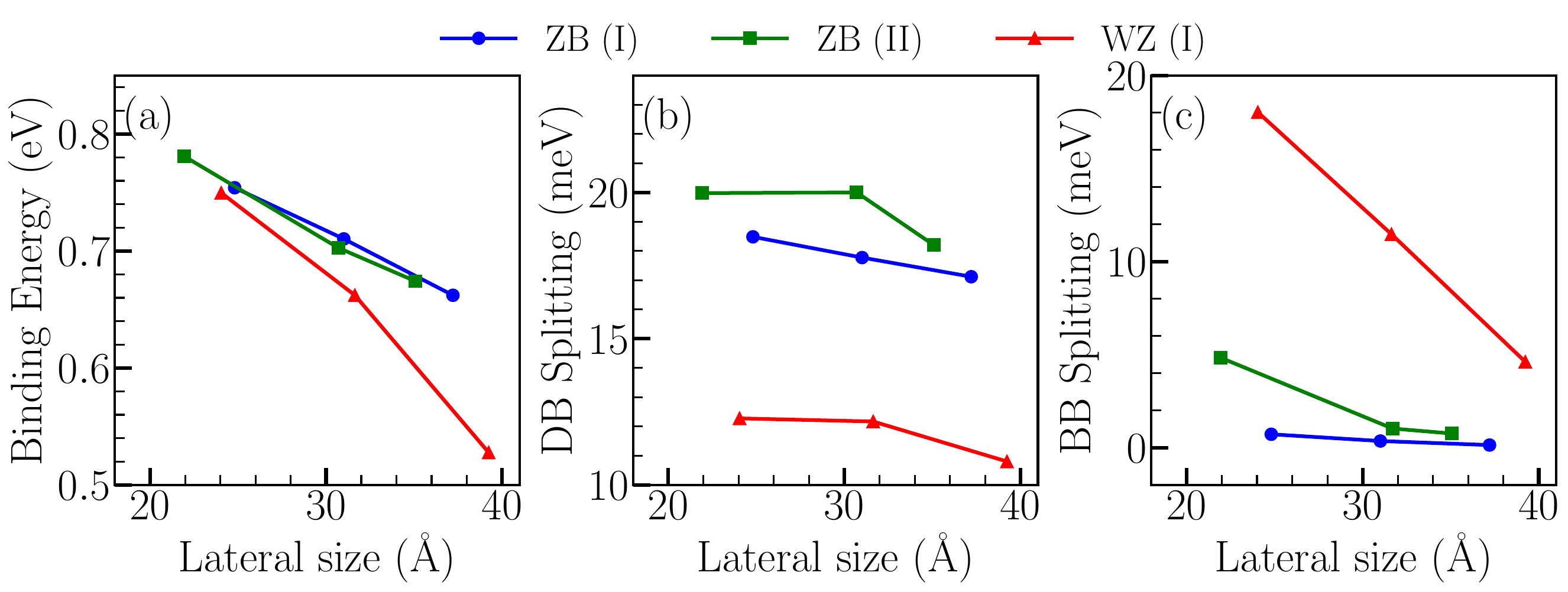}
    \caption{ (a) Exciton Binding energy (b) dark-bright (DB)  splitting, and  (c) bright-bright (BB) splitting in CdSe nanostructures as a function of lateral size. The data includes results for ZB (I), ZB (II), and WZ (I) geometries.
 } \label{BE_BB_BD_vs_size}
\end{figure}

Figure~\ref{BE_BB_BD_vs_size} (a) presents the computed exciton binding energies as a function of nanoplatelet size. The exciton binding energy corresponds to the energy required to dissociate an exciton into a free electron and hole and is a direct measure of the strength of the Coulomb interaction between the carriers. As expected, the binding energy decreases with increasing lateral size for all geometries, due to reduced quantum confinement.  
Experimentally, CdSe nanoplatelets with lateral dimensions of roughly 15–30 nm × 8–20 nm exhibit exciton binding energies ranging from 430 meV (three monolayers) to 264 meV (seven monolayers)~\cite{ji2020dielectric}, consistent with our calculations.


Our computational study, performed for smaller lateral sizes 20–40\,\text{\AA}, naturally yields larger binding energies, as expected from stronger quantum confinement in these smaller platelets.

Figure~\ref{BE_BB_BD_vs_size}(b) shows the dark-bright (DB) exciton splitting as a function of lateral size for the three geometries: ZB(I), ZB(II), and WZ. The dark-bright splitting reflects the energy difference between the lowest bright and the nearly-twofold dark exciton states (see Fig.~\ref{3case_FSS_compare01}), which arises primarily due to the electron-hole exchange interactions. As the lateral size increases, the DB splitting generally decreases across all geometries, indicating reduced quantum confinement and reduced exchange interaction strengths. Experiments report dark-bright exciton splittings of 3–6 meV for CdSe nanoplatelets with thicknesses of 3–5 monolayers~\cite{shornikova2018addressing}. Because the exchange interaction that drives this splitting scales inversely with the spatial extent of the exciton, the splitting decreases as the lateral size increases. Our calculated values of 12–20 meV for much smaller lateral platelets therefore fall in a consistent size-dependent trend.

Figure~\ref{BE_BB_BD_vs_size}(c) presents the bright-bright (BB) exciton splitting, which represents the energy separation between the two lowest optically active (bright) exciton states (see Fig.~\ref{3case_FSS_compare01}). This splitting is the result of electron-hole exchange interaction as well, but influenced by the in-plane asymmetry of the confinement potential. This potential asymmetry leads to two slightly different values of the electron-hole exchange interaction and a lifting of the degeneracy of the two bright exciton states\cite{bester2003pseudopotential}. 
Experimental studies have observed bright-bright splittings of 1-3~meV in colloidal CdSe WZ quantum dots~\cite{Furis2006PRB,Htoon2008PRB}, and similar splittings have also been reported in other nanoplatelets, including zincblende CdSe/CdTe (50 $\mu$eV) and Cs$_{n-1}$Pb$_n$Br$_{3n+1}$ perovskites (5-16~meV)~\cite{CdSeCdTeNPL,PerovskiteNPL}.

In an idealized zincblende nanoplatelet with perfectly square lateral dimensions and symmetry-equivalent edges, the two in-plane bright states would be degenerate, with zero BB splitting. However, real atomistic structures deviate from this ideal picture. In the three-layer Cd-terminated ZB(I) structure, enforcing Cd termination on both surfaces requires local adjustments of atomic positions near the corners, which slightly breaks the in-plane symmetry and produces a small residual splitting of 720, 358, and 142 \textmu eV for structures with lateral widths of 24.81, 31.01, and 37.21 \text{\AA}, respectively. For the ZB(II) geometry, while the top and bottom planes are equivalent and the sides contain a mix of Cd and Se, the corners are not always symmetric, and this asymmetry influences the lateral confinement environment, resulting in somewhat larger BB splittings of 4.8, 1.0, 0.8 meV for structures with lateral widths of 21.93, 31.70, 35.08 \text{\AA}, respectively), which are somewhat larger than those found in ZB(I). The WZ structure intrinsically lacks fourfold rotational symmetry in the basal plane and therefore shows the largest BB splitting among the three cases, with values of 18.0, 11.4, and 6.0 meV for structures with lateral widths of 24.05, 31.64, and 39.24  \text{\AA}, respectively.. For all geometries, the BB splitting decreases with increasing lateral size, consistent with the reduced sensitivity of the wavefunction to the edges of the NPLs.

\section{Conclusions}
In summary, we have presented an accurate and computationally efficient methodology to calculate electron-hole Coulomb interactions in quasi-two-dimensional nanoplatelets, incorporating a 2D screening function and a truncated Coulomb potential specifically adapted to the geometry of these systems. We then introduced a combination of first-principles calculations and many-body treatment: single-particle states were obtained from DFT, followed by screened configuration interaction 
calculations to capture excitonic effects.
We applied this framework to three structural geometries of CdSe nanoplatelets: Zincblende (I), Zincblende (II), and Wurtzite (I) and investigated the exciton binding energy, as well as the exciton fine structure characterized by the dark-bright (DB) and the bright-bright (BB) splittings.
The ZB structures exhibit relatively small, but non-vanishing, BB splitting  (0.1-4.8 meV)  that we attribute to atomistic edge and corners effects slightly breaking the symmetry. The Wurtzite structures show a significantly larger BB splitting, attributed mainly to the intrinsic lower crystallographic symmetry. 
Our results demonstrate clear trends across all geometries: all excitonic features decrease with increasing lateral size, due to reduced confinement but at different rates. These results not only help us better understand the excitonic behavior in different geometries of CdSe nanoplatelets, but also establish a transferable and efficient computational strategy for modeling excitons in other quasi-2D materials.

\section*{Acknowledgments}
This work was supported by the Cluster of Excellence “CUI: Advanced Imaging of Matter”  of the Deutsche Forschungsgemeinschaft (DFG) — EXC 2056 — project ID 390715994.  Computational resources were provided by the Hummel-2 HPC cluster at the University of Hamburg Regional Computing Center (RRZ).

\bibliography{ref}

\end{document}